# The Fornax3D Survey — A Magnitude-Limited Study of Galaxies in the Fornax Cluster with MUSE

Marc Sarzi[1]
Enrichetta Iodice[2]
on behalf of the Fornax3D collaboration

[1] Armagh Observatory & Planetarium, UK
[2] INAF–Astronomical Observatory of Capodimonte, Napoli, Italy

**The Fornax galaxy cluster is an ideal nearby laboratory in which to study the impact of dense environments on the evolution of galaxies. The Fornax3D survey offers extended and deep integral-field spectroscopic observations for the brightest 33 galaxies within of virial radius of the Fornax cluster, obtained with the MUSE integral-field spectrograph, mounted on Unit Telescope 4 (Yepun) of ESO's Very Large Telescope in Chile. The Fornax3D data allowed us to reconstruct the formation of early-type galaxies in the cluster and to explore the link with spiral galaxies. Results have been published in 19 refereed papers since 2018. In this paper we review the broad goals of this campaign, its main results and the potential for future studies combining the MUSE data with the abundant multi-wavelength data coverage for Fornax.**

## Background and goals of the Fornax3D survey

Galactic environment is one of the key drivers of galaxy evolution, with galaxies losing their ability to form new stars and gradually transforming from late-type to early-type objects as they transition to dense galaxy groups or clusters (for example, Dressler, 1980). Various mechanisms have been proposed to explain such a morphological transformation, from gravitational perturbations with other group or cluster members (for example, Moore et al., 1996) to hydrodynamic interactions with the hot and dense gas medium that permeates high-density regions (for example, Boselli & Gavazzi, 2006). The use of integral-field spectroscopy (IFS) has led to significant advances in our understanding of the environmental evolution of galaxies, directly as it happens for instance in ram-pressure stripped objects (for example, Poggianti et al., 2017) and indirectly through the detailed study of the structure of nearby early-type galaxies (ETGs) to infer how their morphological transition occurred, leading to their present kinematic divide between slow- and fast-rotating objects (Emsellem et al., 2007, 2011).

With its extended spectral range, fine spatial sampling, large field of view, and superb throughput, the MUSE integral-field spectrograph (Bacon et al., 2010) is a unique instrument with which to address these (and related) outstanding issues, as demonstrated by some of the first MUSE studies of both slow- and fast-rotating ETGs (Emsellem, Krajnovic & Sarzi, 2014; Guerou et al., 2016). This is particularly the case when MUSE data are combined with sophisticated orbit-superposition models (for example, Krajnovic et al., 2015), which already showed how fast-rotators indeed also contain a dynamically warm component resembling the thick disc of spiral galaxies (Zhu et al., 2018). At the same time MUSE has the potential to further explore the stellar population properties of stellar halos, following in the footsteps of earlier IFS investigations (Weijmans et al., 2009) and allowing a connection to be made with the increasing imaging constraints (for example, Iodice et al., 2016) and modelling predictions (for example, Cook et al., 2016). In this context, the Fornax cluster is an ideal laboratory in which to study the archaeological record of environmental processes using MUSE observations, thanks to its proximity (around 20 Mpc; Blakeslee et al., 2009), more characteristic intermediate cluster mass ($M_{vir} < 10^{14}\ M_{\odot}$; Drinkwater, Gregg & Colless, 2001), southern sky location and plethora of available ancillary data, ranging from optical Hubble Space Telescope (Jordán et al., 2007) and deep Very Large Telescope (VLT) Survey Telescope multi-band images from the Fornax Deep survey (Iodice et al., 2016) to millimetre-wavelength and radio data from the Atacama Large Millimeter/submillimeter Array (ALMA) and MeerKAT (Zabel et al., 2019; Serra et al., 2016) mapping the molecular and HI gas reservoirs of Fornax galaxies.

These considerations motivated the Fornax3D survey (Sarzi et al., 2018; Iodice at al., 2019b), which obtained deep and extended MUSE observations for all 33 galaxies within the virial radius of the Fornax cluster with an apparent *B*-band magnitude brighter than 15. The Fornax3D observations, totalling 106 hours of VLT time, reached a spectral signal-to-noise (S/N) ratio of 25 out to a *B*-band surface brightness of 25 mag. arcsec$^{-2}$ after binning and a S/N of 100 within each central MUSE pointing with little or no binning. This allowed us to pursue the following main objectives: a) fully characterise the embedded discs of ETGs using both orbit-superposition dynamical models and high-quality stellar-population measurements; b) map in detail the stellar-population properties of ETGs, including constraining the slope of the low-mass end of the stellar initial mass function (IMF) and obtaining age and metallicity gradients for their stellar halos; c) provide a census of unresolved sources such a globular clusters and planetary nebulae; d) assess the impact of environmental effects on Fornax LTGs tapping also on ancillary data; and e) provide a rich dataset for the astronomy community.

## The assembly history of the Fornax cluster

The increasing number of cosmological simulations dealing with the formation and evolution of galaxies has in recent years allowed the evolution of significant samples of simulated high-density environments to be followed (for example, Pillepich et al., 2018). Iodice et al. (2019b) improved the picture of the structure and assembly history of the Fornax cluster by combining measurements of the photometric, stellar-population and star-formation properties of Fornax galaxies from the Fornax Deep Survey (FDS; Iodice et al., 2016) images and the Fornax3D MUSE data, placing them in the context of numerical predictions of the phase-space distribution of galaxies in Fornax-like clusters.

This led to the identification of three well-defined structures in Fornax: the core, the north-south (NS) clump and the infalling galaxies (Figure 1, left panel). The core is dominated by the potential of the brightest cluster galaxy, NGC 1399, which coincides with the peak of the X-ray emission. The NS clump galaxies are the reddest, most massive and metal-rich galaxies in the sample, and few of them show ionised gas emission. The infalling galax-ies are distributed nearly symmetrically around the core, at larger projected distance than the galaxies in the clump. Most infalling galaxies are late-type galaxies (LTGs) with ongoing star formation, signs of interaction, a disturbed molecular gas (Zabel et al., 2019; Raj et al., 2019) and on average a lower stellar metallicity compared to objects in the NS clump.

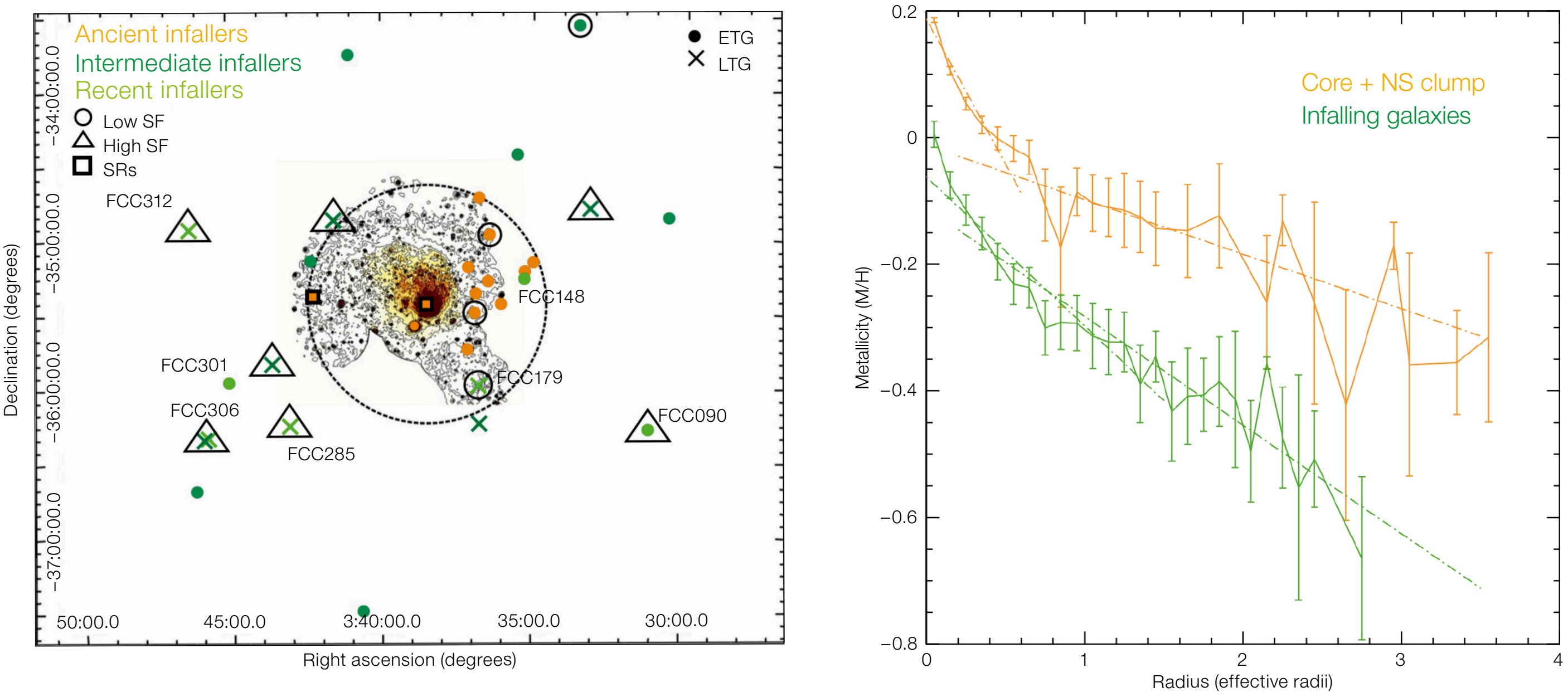

Figure 1. Left panel: Distribution of the Fornax3D sample ETGs (circles) and LTGs (crosses) on the sky. The right ascension and declination (J2000.0) are given in degrees on the horizontal and vertical axes, respectively. The background image and contours map the X-ray emission in the energy range 0.4–1.3 KeV as measured by XMM-Newton (Frank et al., 2013). The dashed circle indicates the transition from the high-density region to the low-density region of the cluster at 0.4 $R_{vir}$ ~ 0.3 Mpc. Orange, green, and light-green symbols represent galaxies identified in phase-space as ancient, intermediate, and recent infallers, respectively, with the last of those being also individually labelled. Black open triangles and circles point to galaxies with high and low star formation activity, respectively, with open black squares further showing the only two slow-rotator ETGs in the Fornax3D sample. Right panel: Running mean as a function of radius of the azimuthally-averaged metallicity for the sample galaxies in the core-NS clump (orange symbols) and for those infalling in the Fornax cluster (green symbols), as also shown in the left panel. The dash-dotted lines represent the fits to the metallicity gradients between transition radii from the bounded in-situ inner component to the accreted ex-situ stellar halo (Spavone et al., 2020).

Comparing the phase-space position of the Fornax3D objects to that of simulated galaxies (Rhee et al., 2017) confirms that the infalling galaxies are either intermediate or recent infallers that entered the cluster less than 4 Gyr ago. Conversely, all but one of the NS clump galaxies fall in the simulated loci of ancient infallers, indicating that the clump may have resulted from the gradual accretion of a group of galaxies more than 8 Gyr ago, most likely along the cosmic web filament connecting the Fornax-Eridanus large-scale structure.

## Pushing into the halos of early-type galaxies

The phase-space segregation of galaxies in the Fornax cluster is also mirrored by systematic differences in the properties of their stellar outskirts. Taking advantage of the extended coverage and depth of the Fornax3D data, Spavone et al. (2022) derived the azimuthally-averaged radial profiles of the stellar velocity dispersion, inclination-corrected specific angular momentum, age, and metallicity for the non-central ETGs of the Fornax3D sample, out to distances of ~ 2–3 effective radii and well beyond the first transition radius, from the bounded in-situ component to the accreted (ex-situ) stellar halo, as previously derived by Spavone et al. (2020) using deep FDS images.

As shown in Figure 1 (right panel), galaxies in the core and NS clump of the cluster, which have the highest accreted mass fraction, show milder metallicity gradients in their outskirts than the galaxies falling into the cluster. This difference in outer metallicity gradients and accreted mass fraction between the galaxies in the two main Fornax sub-structures reinforces the idea that the NS clump may result from the accretion of a group of galaxies during the gradual build-up of the cluster, while the infalling galaxies entered the cluster later. Pre-processing mechanisms in the clump, such as repeated mergers, may indeed have been responsible for shaping the stellar halo around such galaxies, feeding this component in terms of baryonic mass and producing a mixing of different stellar populations from the accreted satellites that resulted in a flatter metallicity radial profile at larger radii.
On the other hand, the lack of an extended stellar envelope in the infalling galaxies is consistent with their steeper metallicity gradients.

## State-of-the-art mapping of stellar population properties

The Fornax3D MUSE data cover an extended area of galactic discs in Fornax, on average out to two half-light radii, allowing our understanding of the formation of discs in ETGs to be tested.
In particular, high-quality two-pointing mosaics of the three edge-on lenticular galaxies in the Fornax3D sample pro-vided a unique view of their vertical structure (Pinna et al., 2019a,b). In all three objects our stellar-population maps indicated the presence of an old, metal-poor thick disc with enhanced alpha-element abundances and of a more metal-rich but less alpha-enhanced thin disc. One galaxy, located close to the cluster centre, shows overall very old populations, probably due to a strong impact of star-formation quenching processes in both cluster and pre-cluster environments. On the other hand, the position of the other two galaxies in less dense regions of the cluster may have allowed a more prolonged star formation, resulting in younger ages for their thin disc. Spatially resolved star-formation histories for these edge-on objects also revealed the presence of younger and more metal-poor sub-populations in their thick discs, probably accreted during past mergers.

A more complete mapping of the stellar age, metallicity and alpha-element abundance across the sample was provided by Martín-Navarro et al. (2019, 2021), who exploited the exquisite quality of Fornax3D data to also map, for the first time, the two-dimensional variations of the IMF beyond the Milky Way (Figure 2). The IMF maps revealed a striking surprise: metallicity alone is not able to explain the complex behaviour of IMF variations, which appear also to be coupled to the internal orbital structure of galaxies (Poci et al., 2022).

## Chemo-dynamical modelling dissection of early-type galaxies

Building on previous modelling efforts to dissect the stellar-population components of ETGs (for example, Zhu et al., 2018), we have developed new orbit-superposition techniques capable of matching simultaneously not only the observed stellar surface brightness and kinematics but also maps of the stellar age and metallicity as shown in Figure 2 (Poci et al., 2019; Zhu et al., 2020). By applying this to all 23 ETGs in the Fornax3D survey we obtain their internal stellar orbit distributions as well as their age and metallicity distribution, which we can use to separate different components in a physical and flexible way. Figure 3 (top panels) illustrates the case of FCC167, where we isolate a dynamically cold, metal-rich and relatively younger disc, a concentrated dynamically hot and metal-rich bulge, a more extended and metal-poor inner stellar halo and a dynamically warm component.

Using these models we derived several interesting results: (1) for the three edge-on galaxies in Fornax we derived the stellar age-velocity dispersion profile of their discs and found evidence that metallicity may be a key driver of this relation (Poci et al., 2021), making them directly comparable with high-redshift galaxies; (2) in the massive ETGs FCC 167 and FCC 276, we ascribed the formation of their inner stellar halo to a merger with a now-destroyed massive satellite (Zhu et al., 2022a,b), making it possible to in future quantify the timing and accreted mass during ancient massive mergers for large samples of nearby galaxies; and (3) we isolated the dynamically-cold disc of all Fornax3D ETGs and found that the galaxies that fell into the cluster early on have significantly lower cold-disc mass fractions than the recently infalling galaxies (Figure 3, lower panels) and that cold discs in ETGs have positive age gradients, supporting an outside–in quenching of their star formation while falling into the cluster (Ding et al., 2022, submitted to *Astronomy & Astrophysics*).

## Zooming-in on globular clusters and planetary nebulae

Because of their old ages, globular clusters (GCs) are regarded as important fossil tracers of galaxy evolution. Fahrion et al. (2020a) used the excellent image quality of the Fornax3D data, with a median FWHM of 0.9 arcseconds, to extract and analyse the spectra of GCs even in the central regions of the Fornax3D sample galaxies. MUSE allowed us to accurately account for the background in each GC spectrum, collecting an extensive spectroscopic GC catalogue with 722 GC velocities and 238 metallicity measurements. Using this data set, we explored how well GCs trace the underlying galaxy properties and found GCs to be valuable tracers of the enclosed mass as well as the galaxy metallicity, from the central regions out to several effective radii. Additionally, we established a nonlinear relation between GC metallicities and photometric colours that has strong implications for the merger histories inferred solely from GC colours (Fahrion et al., 2020b).

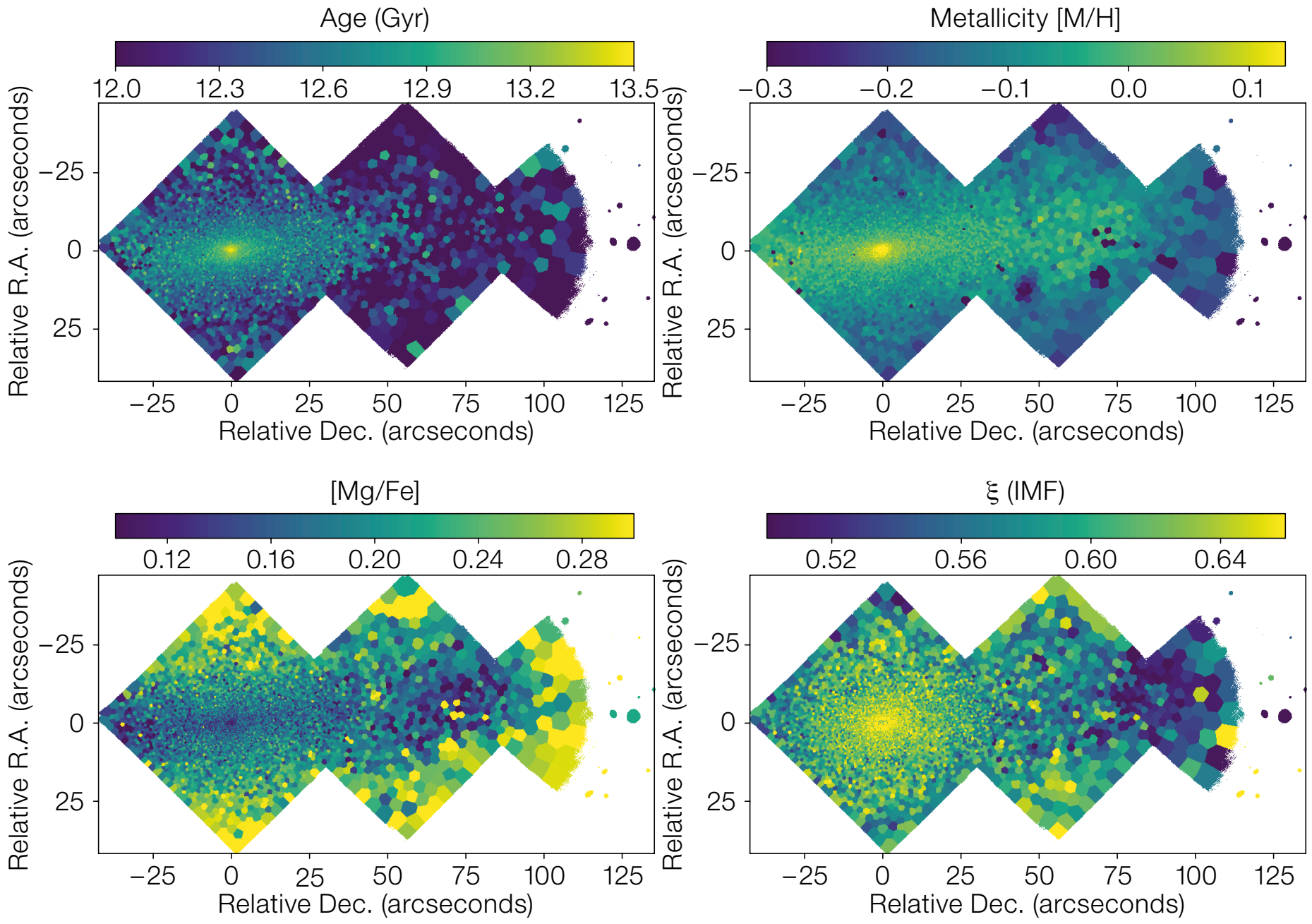

Figure 2. Stellar population maps of NGC 1380 (FCC167). Top left panel: age map where a relatively older central component is clearly visible. The metallicity (top right) and the magnesium abundance [Mg/Fe] (bottom left) maps show the clear signature of a chemically evolved disc, confined within a vertical height of ~10 arcsec, which coincides with the kinematically cold component observed in this galaxy. The stellar dwarf-to-giant ratio map (ξ, bottom right panel), which is tracing the variation of the initial mass function (IMF) over the field of view of the galaxy, exhibits a different 2D structure, however, that does not follow the metallicity variations, but closely follows the dynamically hot and warm stellar component revealed in this galaxy thanks to orbit-superposition models (Sarzi et al., 2018).

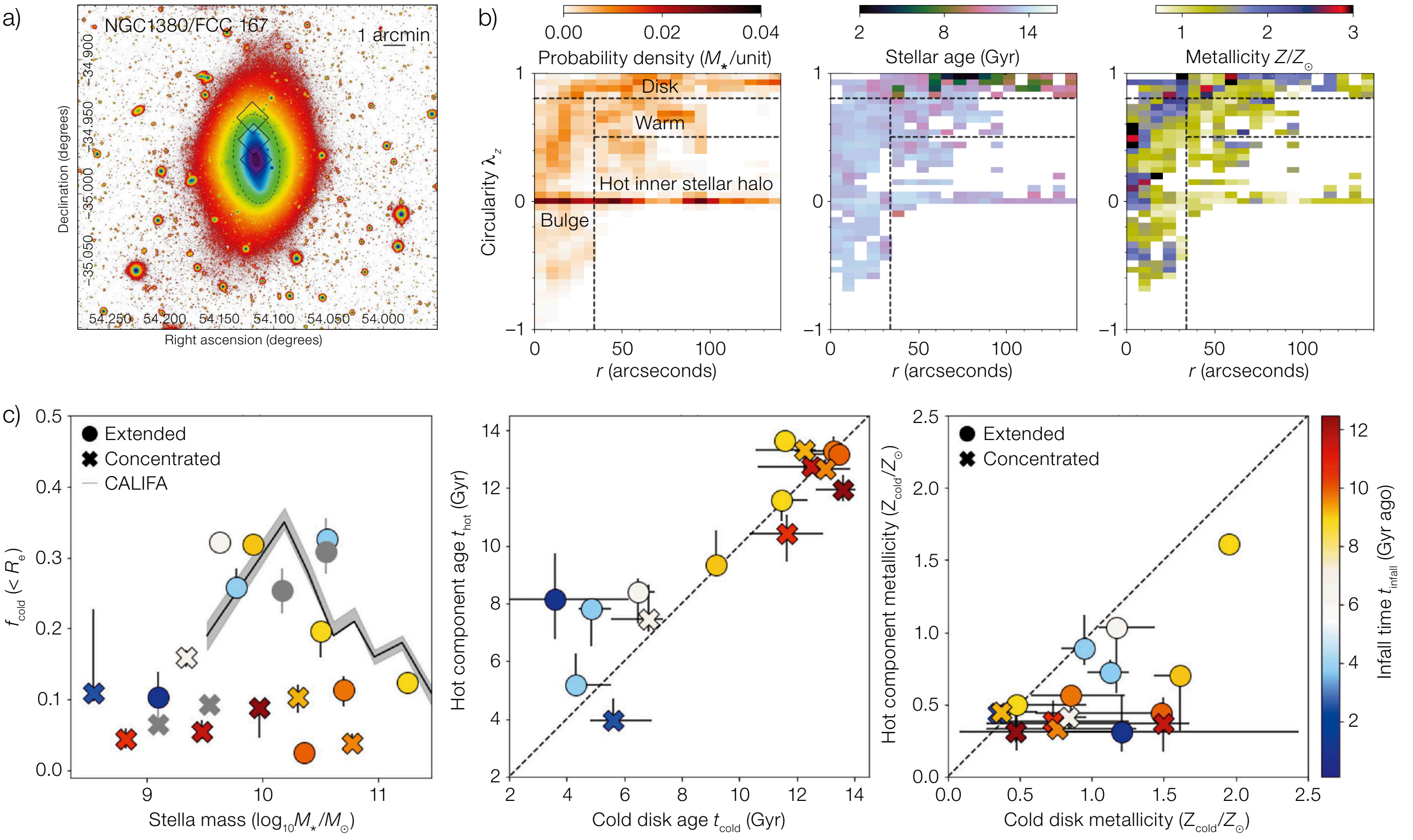

Figure 3. Panel (a): Deep *r*-band image of NGC 1380 (Iodice et al., 2019a). The three diamonds show the position of the Fornax3D MUSE pointings whereas the dashed black ellipse traces the isophote at a *B*-band surface brightness level of 25 mag arcsec$^{-2}$. Panels (b): Population-orbit superposition model results for NGC 1380. The left panel shows the probability density distribution of the stellar orbits p($r$, $\lambda_z$) in the phase space of time-averaged radius $r$ versus circularity $\lambda_z$ in units of stellar mass per unit area in phase space, normalised to the total mass within the coverage of the MUSE data. The central and right panels show the stellar age $T(r, \lambda_z)$ and metallicity $Z(r, \lambda_z)$ distribution of the orbits in the same phase-space, respectively. Panels (c): Properties of the dynamically cold discs in the Fornax3D ETGs. The left panel shows the cold-disc fraction as a function of total stellar mass whereas the central and right panels compare the average stellar age and metallicity of cold discs to that of dynamically hot and warm components. In all three panels, galaxies are colour-coded according to their cluster infall time and shown with different symbols according to whether they are concentrated or extended. Galaxies that entered the cluster more recently tend to have extended and relatively more massive discs, which can also show younger populations. More generally stars in dynamically-cold discs tend to be more metal-rich, across the whole sample.

Planetary Nebulae (PNe) are incredibly luminous sources of [OIII]λ5007 nebular emission that can be found across different galaxies and used to derive the distances to their host galaxies, thanks to the apparently invariant bright-end cut-off of the PNe luminosity function (PNLF; for example, Ciardullo et al., 1989). In this respect, integral-field spectroscopy can secure well-sampled PNLFs by making it possible to probe the central regions of galaxies that are indeed rich in PNe (for example, Sarzi et al., 2011; Kreckel et al., 2017). Spriggs et al. (2020, 2021) used the Fornax3D MUSE data to identify 1350 PNe across 21 ETGs, deriving PNLF distance measurements that agree well with previous distances based on surface brightness fluctuations (Blakeslee et al., 2009). With these individual measurements, we arrive at an average distance to the Fornax cluster itself of 19.86 ± 0.32 Mpc. These are encouraging results, considering the potential of adaptive-optics-assisted MUSE observations to push even further the reach of PNLF distance measurements.

## Combining with ancillary data and the value to the Community

The value of the Fornax3D data increases further when they are used together with ancillary data, especially when studying the evolution of gas and dust in cluster galaxies. For instance, Viaene et al. (2019) combined MUSE measurements for the dust attenuation curve with radiative transfer models based also on Herschel, WISE and SPIRE data to conclude that the central dust disc of NGC1380 lacks small grains. At the same time, dust destruction timescales from sputtering based on Chandra X-ray data indicate that such dust must be strongly self-shielding and clumpy or else it would already have disappeared.

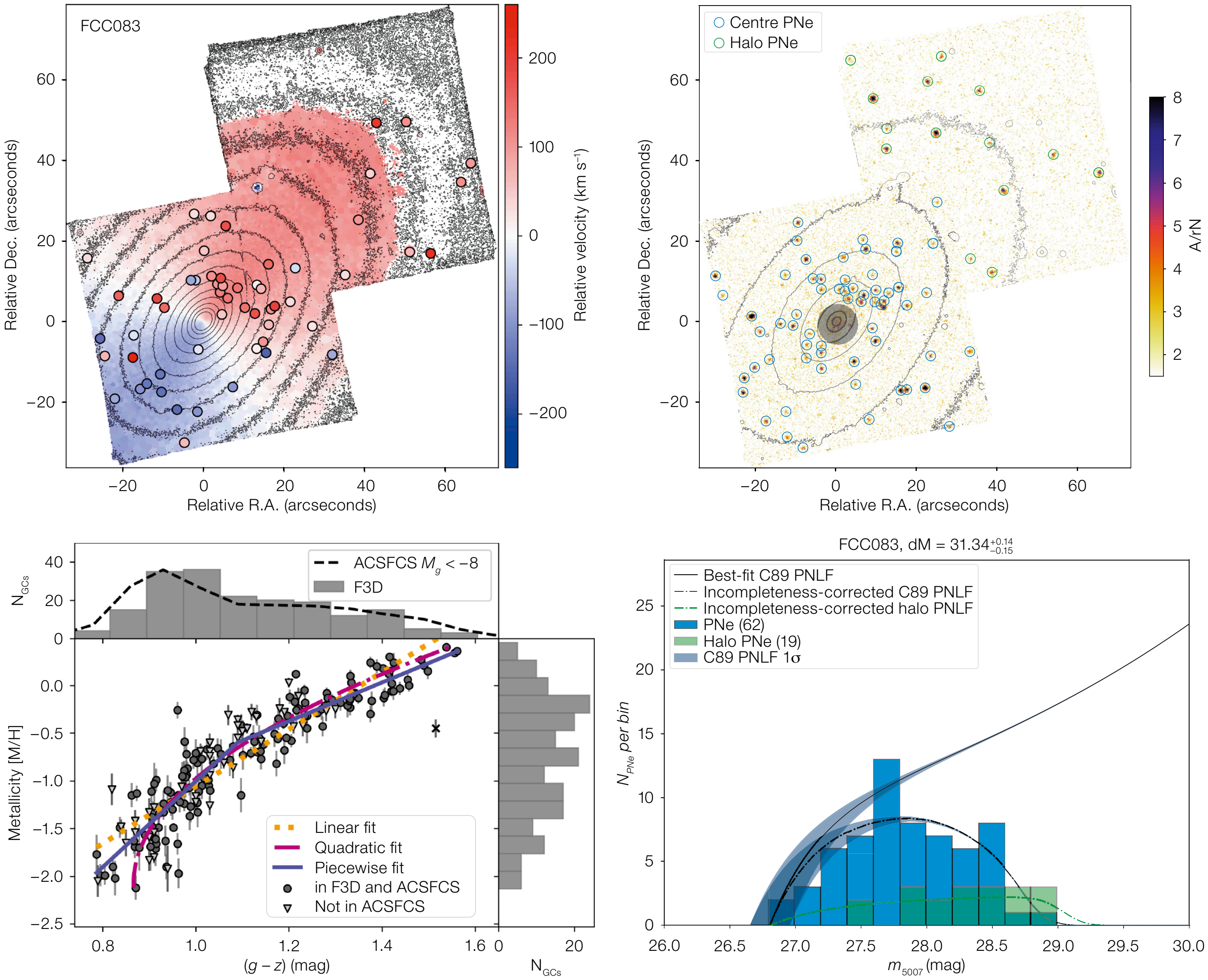

**Figure 4.** GCs and PNe in the Fornax3D galaxies. Left panels: MUSE stellar velocity for FCC083, with GCs detections and their velocities (top) and non-linear colour-metallicity relation of GCs across the entire sample (bottom). Right panels: Map of the the amplitude-to-residual noise ratio (A/rN) between the peak amplitude of the (OIII)λ5007 emission over the noise level in the residuals of our fit to the MUSE spectra showing the location of PNe point sources (top) and resulting PNLF for each MUSE pointing with corresponding fits fully accounting for PNe detection incompleteness and delivering the apparent magnitude of the bright cutoff (bottom).

Using data from the ALMA Fornax Cluster Survey (Zabel et al., 2019) we further explored the depletion timescale for the gas reservoirs in 15 Fornax galaxies by comparing, on the scale of 300 pc, the MUSE extinction-corrected star formation rates with ALMA measurements for the molecular gas content. In this way we found that gas depletion times appear to shorten closer to the cluster centre, albeit that this trend is mainly driven by dwarf galaxies with disturbed molecular gas reservoirs (Zabel et al., 2020). In a second work, we derived the dust-to-gas ratio in Fornax galaxies, computed using far-infrared Herschel observations to measure the total dust mass and data from the Australia Telescope Compact Array (ATCA; Loni et al., 2021) to obtain the total neutral gas mass values in addition to the ALMA molecular gas mass. We compared these total dust-to-gas ratios to mean values for gas-phase metallicity obtained from the Fornax3D

MUSE data (see also Lara-López et al. [2022] for a spatially resolved study of the gas metallicity in such objects). We find that gas-to-dust ratios in Fornax galaxies are systematically lower than those in field galaxies at fixed metallicity. This implies that a relatively large fraction of the metals in these Fornax systems is locked up in dust, which is possibly due to altered chemical evolution as a result of the dense environment (Zabel et al., 2021).

To conclude, the Fornax3D data also have a legacy value that is already driving progress in unanticipated ways. A case in point is represented by the work of Smith (2020) who detected seven novae via the analysis of individual Fornax3D exposures taken several months apart.

## Data Release

The data release of the Fornax3D survey through the ESO Science Archive Facility is foreseen for the first quartile of 2023. It will contain reduced and calibrated mosaics for all galaxies and will be announced in the ESO Science Newsletter.

ESO/P. Horálek

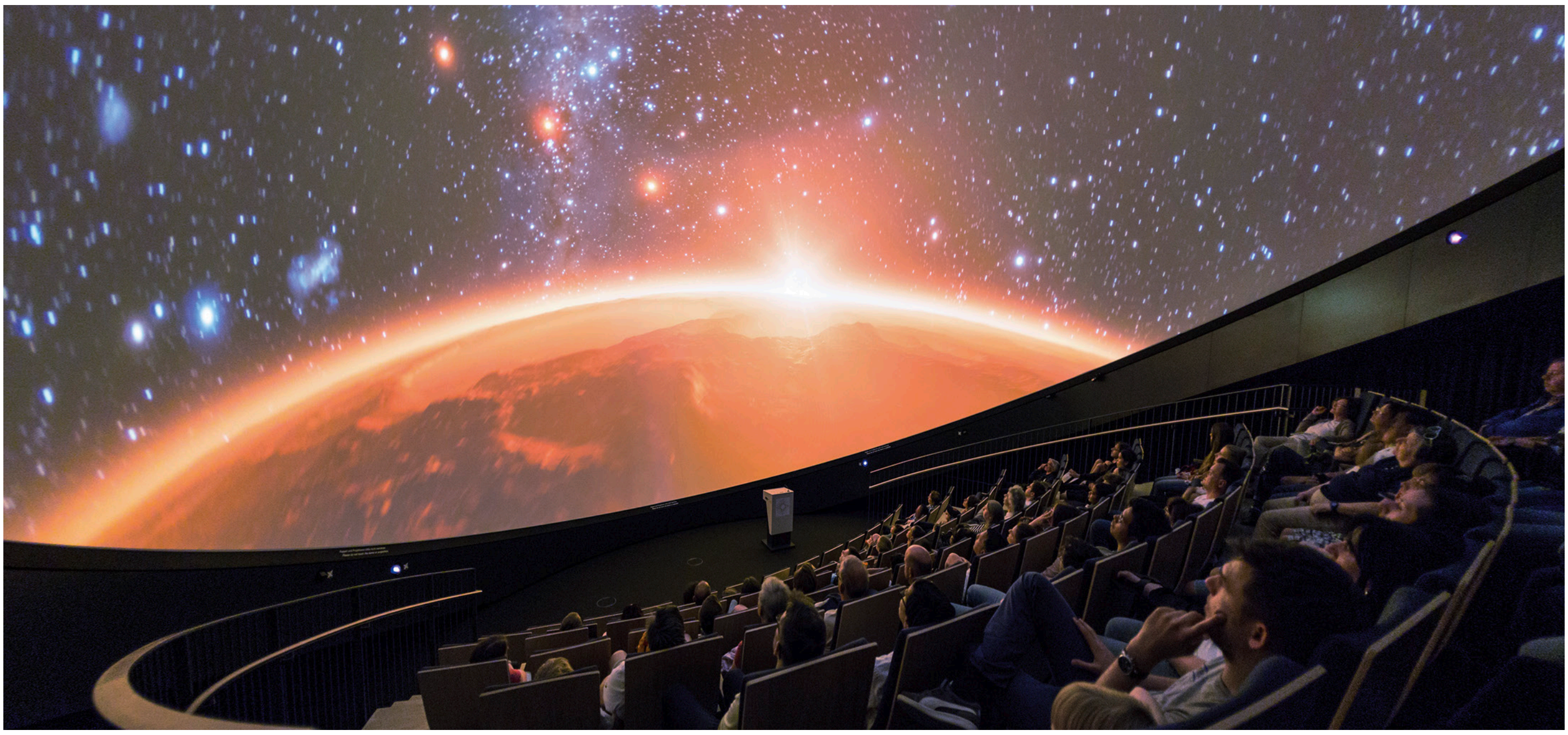

A view from inside the planetarium at the ESO Supernova Planetarium & Visitor Centre, which opened its doors to the public on Saturday 28 April 2018. The building is open five days a week and features planetarium screenings, tours and a permanent exhibition in both German and English. The 25-degree tilted planetarium dome does not just give the audience the sensation of watching the Universe, but of being immersed in it.